\documentclass{extarticle}
\usepackage{amsfonts,amssymb,amsmath,amsthm}
\usepackage{url}
\usepackage{enumerate}
\usepackage{color}
\usepackage{nameref}
\usepackage[dvipsnames]{xcolor}
\usepackage[colorlinks=true,linkcolor=RoyalBlue,citecolor=PineGreen,urlcolor=blue]{hyperref}
\usepackage{comment}
\usepackage{enumitem}
\usepackage[margin=3cm,letterpaper]{geometry}
\usepackage{bm}
\usepackage{times}
\usepackage{natbib}

\title{SIAN: software for structural identifiability analysis of ODE~models}

\author{Hoon Hong\footnote{{hong@ncsu.edu}, Department of Mathematics, North Carolina State University, Raleigh, USA}, Alexey Ovchinnikov\footnote{{aovchinnikov@qc.cuny.edu}, Department of Mathematics, CUNY Queens College and Ph.D. Programs in Mathematics and Computer Science, CUNY Graudate Center, New York, USA}, Gleb Pogudin\footnote{{pogudin@cims.nyu.edu}, Courant Institute of Mathematical Sciences, New York University}, and Chee Yap\footnote{{yap@cs.nyu.edu}, Courant Institute of Mathematical Sciences, New York University}}
\date{}

\begin{document}

\maketitle

\begin{abstract}
Biological processes are often modeled by ordinary differential equations with unknown parameters. 
The unknown parameters are usually estimated from experimental data.
In some cases, due to the structure of the model, this estimation problem does not have a unique solution even in the case of continuous noise-free data.
It is therefore
desirable to check the uniqueness a priori 
before carrying out actual experiment. 
We present a new software SIAN (Structural Identifiability
ANalyser) that does this.
Our software can tackle problems that could not be tackled by previously  developed
packages.
\end{abstract}

\section{Introduction}

Ordinary differential equations (ODEs) with unknown parameters are widely used
for modeling biological processes and phenomena. One is often interested in the values of these parameters due to their importance, 
as, e.g., they may represent key biological mechanisms or targets for intervention.
A standard way to find the values of the parameters from 
experimental data is to find the parameter values that fit 
the data with minimal error, typically framed from a statistical perspective as maximum likelihood or Bayesian inference.

However, it might happen that, due to the structure of the model, it is impossible to recover the value of a parameter of interest from the data even assuming the ideal case of a continuous noise-free data.
If this is the case, then regardless of the chosen data fitting approach, it is impossible to guarantee that it will find the correct parameter value.
As we will see, this structural property can be assessed \emph{a priori} without conducting (often costly) experiments.
Thus, a crucial first step to any parameter estimation problem is to check whether the parameter of interest is \emph{structurally globally identifiable}, i.e., the parameter value can be recovered uniquely from the data under the assumption that the data is continuous and noise-free.
We explain the notion of global identifiability in more detail 
 in Section~\ref{sec:features}. 
 For a formal definition and illustrating examples, we refer to~\citep[Section~2]{HOPY}.

We present SIAN (Structural Identifiability ANalyser), our new software for assessing identifiability for ODE models,
based on the algorithm developed and rigorously justified in
~\citep{HOPY}.

%%%%%%%%%%%%%%%%%%%%%%%%%%%%%%%%%%%%%%

\section{Existing software for structural identifiability}
\label{sec:stateofart}

Assessing global identifiability is a challenging problem.
Hence a weaker notion called ``local identifiability'' was introduced and tackled first.
``Local'' indicates that a parameter can be identified locally (in some neighborhood).
For a polynomial system, it is the same as saying that a parameter can be identified up to finitely many options.
There are fast and reliable software packages for assessing local identifiability such as ObservabilityTest~\citep{Sed2002} and EAR~\citep{EAR}.

However, even relatively simple real-life systems
can involve 
locally but not globally identifiable parameters
(see~\cite[Section~4]{DAISY2018}, \citep{Nor1982}, and Supplementary Materials~A.1).
Thus, it is highly desirable to have software that could assess global identifiability.
There has been significant progress in this direction:
\begin{itemize}
    \item packages DAISY~\citep{DAISY} and COMBOS~\citep{COMBOS} are based on the approach via input-output equations and can check global identifiability for systems with the ``solvability'' property (see~\citep[Example~6]{HOPY} for a discussion).
    \item GenSSI~2.0 package~\citep{LFCBBCH} is based on the generating series approach and checks
    global identifiability conditionally on %an 
    extra input, the 
    truncation order (for a discussion how the 
    truncation order affects the output of the algorithm, see~\citep[Example~7]{HOPY}).
\end{itemize}

%%%%%%%%%%%%%%%%%%%%%%%%%%%%%%%%%%%%%%%%%%

\section{Features}
\label{sec:features}

We present SIAN, software written in {\sc{Maple}}, that has the following input-output specification.

\paragraph{Input.} 
A \textbf{system} $\Sigma$ of the form
\begin{equation}\label{eq:main}
\begin{cases}
    \bm{\dot{x}} = \bm{f}(\bm{x}, \bm{\mu}, \bm{u}),\\
    \bm{y} = \bm{g}(\bm{x}, \bm{\mu}, \bm{u}),\\
    \bm{x}(0) = \bm{x}^\ast,
\end{cases}
\end{equation}
%where
\begin{itemize}
    \item $\bm{x}$ is a vector of state variables,
    \item $\bm{u}$ is a vector of input (control) variables to be chosen by an experimenter,
    \item $\bm{y}$ is a vector of output variables,
    \item $\bm{\mu}$ and $\bm{x}^\ast$ are vectors of unknown scalar parameters and unknown initial conditions, respectively,
    \item $\bm{f}$ and $\bm{g}$ are vectors of rational functions in $\bm{x}$, $\bm{\mu}$, and $\bm{u}$ with complex coefficients (other types of functions can also be handled, see Supplementary Material~A.2)
\end{itemize}
  and \textbf{a real number} $0 < p < 1$, the user-specified probability of correctness of the result. 
  That is, SIAN is a Monte Carlo randomized algorithm, see~\citep[Chapter~1.2]{Randomized}.

\paragraph{Output.} For every $\theta \in \bm{\mu} \cup \bm{x}^\ast$,  
SIAN
assigns one of the following labels:
\begin{itemize}
    \item \textbf{Globally identifiable}: for almost every solution of~\eqref{eq:main}, every solution of~\eqref{eq:main} with the same $\bm{u}$-component and $\bm{y}$-component has the same value of $\theta$.
    \item \textbf{Locally but not globally identifiable}: for almost every solution of~\eqref{eq:main}, among the solutions of~\eqref{eq:main} with the same $\bm{u}$-component and $\bm{y}$-component, there are only finitely many possible values of $\theta$.
    
    \item \textbf{Not identifiable}: for almost every solution of~\eqref{eq:main}, among the solutions of~\eqref{eq:main} with the same $\bm{u}$-component and $\bm{y}$-component, there are infinitely many possible values of $\theta$.
\end{itemize}
The assigned labels are correct with probability at least $p$.

\smallskip

\noindent
We would like to emphasize the following \textbf{extra features}:
\begin{itemize}
    \item SIAN is parallellizable and can take advantage of a multicore computing environment.
    \item SIAN assesses not only the identifiability of the model, but checks individual identifiability of every parameter.
    \item SIAN can assess identifiability of the parameters appearing in the system and the initial condition. Identifiability of initial conditions is often referred to as observability.
\end{itemize}

%%%%%%%%%%%%%%%%%%%%%%%%%%%%%%%%%%%%%%%%%%

\section{Performance and Applications}

In this section, we compare our software with the existing software tools for assessing global identifiability, namely COMBOS, DAISY, and GenSSI (see Section~\ref{sec:stateofart}).
All of the benchmark problems are listed in Supplementary Material B. The source code of the benchmarks problems for COMBOS, DAISY and GenSSI used for the comparison is included into the Supplementary Data.
The source code for the benchmark problems for SIAN is available at
\url{https://github.com/pogudingleb/SIAN/tree/master/examples}.

We use a computer with $96$ CPUs, $2.4$ GHz and CentOS 6.9 (Linux).
The runtimes in Table~\ref{table:runtimes} are the elapsed time.
SIAN was run on {\sc{Maple 2017}} with the probability of correctness $p = 0.99$, GenSSI~2.0 was run on Matlab R2017a, and we used DAISY~1.9.
\begin{table}[h!]
\caption{Runtimes (in minutes) on benchmark problems}\label{table:runtimes}
\begin{center}
\begin{tabular}{|l|r|r|r|r|}
\hline
  Example & GenSSI~2.0 & COMBOS  & DAISY & SIAN \\
\hline
  Chemical Reaction &$\ast$ & $\ast\ast$ & $>$ \hfill $6{,}000$ & $< \hspace{1mm}1$\\
  \hline
  HIV & $>$ \hspace{1mm} $12{,}000$ & $\ast\ast$ & $>$ \hfill $6{,}600$ & $< \hspace{1mm}1$ \\
  \hline
  SIRS w/ forcing & $>$ \hspace{1mm} $12{,}000$ & $\ast\ast$ & $>$ \hfill $6{,}600$ & $< \hspace{1mm}1$ \\
  \hline
  Cholera &$\ast$ & $85\;$ & $30$  & $3$\\
  \hline
  Protein complex & $>$ \hspace{1mm} $12{,}000$ &$\ast\ast$ & $>$ \hfill $6{,}600$ & $47$\\
\hline
  Pharmacokinetics & $>$ \hspace{1mm} $12{,}000$ &$\ast\ast$  & $>$ \hfill $7{,}800$ & $962$\\
\hline
\end{tabular}
\end{center}
\small
$\ast$: \;\;\;\;\;GenSSI 2.0 returns~``{\it Warning: Unable to find explicit solution.}''\\
$\ast\ast$:\;\;\; COMBOS \ returns~``{\it Model may have been entered incorrectly or cannot be solved with COMBOS algorithms.}''
\end{table}

%%%%%%%%%%%%%%%%%%%%%%%%%%%%%%%%

\section*{Acknowledgements}
The authors are grateful to the CCiS at Queens College and CIMS NYU for the computational resources and to Julio Banga, Marisa Eisenberg, Nikki Meshkat, and Maria Pia Saccomani for useful discussions.

\section*{Funding}
This work was supported by the National Science Foundation [CCF-1563942, CCF-1564132, CCF-1319632, DMS-1760448, CCF-1708884]; National Security Agency [\#H98230-18-1-0016]; and City University of New York [PSC-CUNY \#69827-0047,  \#60098-00 48].

\bibliographystyle{abbrvnat}
\bibliography{document}

\end{document}

\maketitle

This document is structured as follows:
\begin{itemize}
    \item Section~\ref{sec:illustrate} contains three illustrating examples mentioned in the text of the paper.
    \item Section~\ref{sec:Benchmarks} contains descriptions of the benchmark problems used in Section~4 of the paper.
\end{itemize}

For the convenience of the reader while navigating between the main paper and the Supplementary materials, we recall that SIAN, software written in {\sc Maple},  has the following input-output specification.

\paragraph{Input.} 
A \textbf{system} $\Sigma$ of the form
\begin{equation}\label{eq:main}
\begin{cases}
    \bm{\dot{x}} = \bm{f}(\bm{x}, \bm{\mu}, \bm{u}),\\
    \bm{y} = \bm{g}(\bm{x}, \bm{\mu}, \bm{u}),\\
    \bm{x}(0) = \bm{x}^\ast,
\end{cases}
\end{equation}
where
\begin{itemize}
    \item $\bm{x}$ is a vector of state variables,
    \item $\bm{u}$ is a vector of input (control) variables to be chosen by an experimenter,
    \item $\bm{y}$ is a vector of output variables,
    \item $\bm{\mu}$ and $\bm{x}^\ast$ are vectors of unknown scalar parameters and unknown initial conditions, respectively,
    \item $\bm{f}$ and $\bm{g}$ are vectors of rational functions in $\bm{x}$, $\bm{\mu}$, and $\bm{u}$ with complex coefficients (other types of functions can also be handled, see Section~\ref{sec:ruminallipolysis})
\end{itemize}
  and \textbf{a real number} $0 < p < 1$, the user-specified probability of correctness of the result.
  That is, SIAN is a Monte Carlo randomized algorithm, see~\citep[Chapter~1.2]{Randomized}.

\paragraph{Output.} For every $\theta \in \bm{\mu} \cup \bm{x}^\ast$, the program  assigns one of the following labels:
\begin{itemize}
    \item \textbf{Globally identifiable}: for almost every solution of~\eqref{eq:main}, every solution of~\eqref{eq:main} with the same $\bm{u}$-component and $\bm{y}$-component has the same value of $\theta$.
    \item \textbf{Locally but not globally identifiable}: for almost every solution of~\eqref{eq:main}, among the solutions of~\eqref{eq:main} with the same $\bm{u}$-component and $\bm{y}$-component, there are only finitely many possible values of $\theta$.
    
    \item \textbf{Not identifiable}: for almost every solution of~\eqref{eq:main}, among the solutions of~\eqref{eq:main} with the same $\bm{u}$-component and $\bm{y}$-component, there are infinitely many possible values of $\theta$.
\end{itemize}
The assigned labels are correct with probability at least $p$.

\section{Illustrating examples}\label{sec:illustrate}

\subsection{Chemical reaction}
\label{sec:slowfast}

\paragraph{Purpose of the example}
\begin{itemize}
    \item to show that locally but not globally identifiable parameters appear even in small systems arising in real-life systems;
    \item to illustrate how one could take into account the possibility of having some of the parameters
    \begin{itemize}
        \item unknown at the stage of creating the model but
        \item become directly known (measured) while performing the experiment.
    \end{itemize}
\end{itemize}

\paragraph{System and discussion.} Consider the following consecutive reaction scheme with three species $A$, $B$, and~$C$:
\[
A \xrightarrow{k_1} B \xrightarrow{k_2} C.
\]
Then the amounts $x_A, x_B$, and $x_C$ of species evolve according to the following system of differential equations
\begin{equation}\label{eq:chem}
\begin{cases}
  \dot{x}_A = -k_1 x_A,\\
  \dot{x}_B = k_1 x_A - k_2 x_B,\\
  \dot{x}_C = k_2 x_B.
\end{cases}
\end{equation}
We assume that, in the experiment, we can observe the amount $x_C$ and a combination $\epsilon_A x_A + \epsilon_B x_B + \epsilon_C x_C$ (where $\epsilon_A$, $\epsilon_B$, and $\epsilon_C$ are parameters), which may represent absorbance, conductivity, or ligand release~\cite[p. 701]{VR1988}.
This gives two outputs $y_1 = x_C$ and $y_2 = \epsilon_A x_A + \epsilon_B x_B + \epsilon_C x_C$.

In addition to this, we are also given~\cite[p.~701]{VR1988} that the values of the parameters $\epsilon_A$ and $\epsilon_C$
will become known at the experiment stage but are unknown at the modeling stage.
We can encode this
within our framework by considering $\epsilon_A$ and $\epsilon_C$ as \emph{observable functions (outputs) with zero derivative}.
In total, we arrive at the following system:
\begin{equation}\label{eq:chem_main}
\begin{cases}
  \dot{x}_A = -k_1 x_A,\\
  \dot{x}_B = k_1 x_A - k_2 x_B,\\
  \dot{x}_C = k_2 x_B,\\
  \dot{\epsilon}_A = 0,\\
  \dot{\epsilon}_C = 0,\\
  y_1 = x_C,\\
  y_2 = \epsilon_A x_A + \epsilon_B x_B + \epsilon_C x_C,\\
  y_3 = \epsilon_A,\\
  y_4 = \epsilon_C,
\end{cases}
\end{equation}
where $\bm{x} = (x_A, x_B, x_C, \varepsilon_A, \varepsilon_C)$, $\bm{y} = (y_1, y_2, y_3, y_4)$, $\bm{\mu} = (k_1, k_2, \epsilon_B)$, and $\bm{x}^\ast = (x_A^\ast, x_B^\ast, x_C^\ast, \varepsilon_A^\ast, \varepsilon_C^\ast)$.

\paragraph{Results} Our software outputs that all the parameters $\bm{\mu}$ and initial values $\bm{x}^\ast$ are locally identifiable, but only $x_C^\ast, \epsilon_A^\ast$, and $\epsilon_C^\ast$ are globally identifiable.
In fact, one can show that the set $\{k_1, k_2\}$ can be always found but any of these two numbers can be either $k_1$ or $k_2$~\cite[Equation~(1.3)]{VR1988}.
In the literature, this phenomenon is referred to as slow-fast ambiguity.

\paragraph{Source code:} \url{https://github.com/pogudingleb/SIAN/blob/master/examples/SlowFast.mpl}.

\paragraph{Remark} Applying SIAN to 
\begin{equation*}
\begin{cases}
  \dot{x}_A = -k_1 x_A,\\
  \dot{x}_B = k_1 x_A - k_2 x_B,\\
  \dot{x}_C = k_2 x_B,\\
  y_1 = x_C,\\
  y_2 = \epsilon_A x_A + \epsilon_B x_B + \epsilon_C x_C,
\end{cases}
\end{equation*}
one can show that the assumption that $\epsilon_A$ and $\epsilon_C$ can be measured separately is redundant: they both are globally identifiable even just from $y_1$ and $y_2$.

%%%%%

\subsection{Ruminal lipolysis}\label{sec:ruminallipolysis}

\paragraph{Purpose of the example} is to show how one can handle the case 
    in which the right-hand side of some of the equations is not a rational function of the parameters.

\paragraph{System and discussion.} The following model of ruminal lipolysis was considered
in~\cite[Equations~(1-5)]{lipolysis}, and its identifiability was discussed in~\cite[Supplementary Material S2]{animals}.

\begin{equation}\label{eq:lip_orig}
\begin{cases}
    \dot{x}_1 = -\frac{k_1 x_1}{k_2  + x_1} e^{-k_3 t},\\
    \dot{x}_2 = \frac{2 k_1 x_1}{3(k_2 + x_1)} e^{-k_3 t} - k_4 x_2,\\
    \dot{x}_3 = \frac{1}{2} k_4 x_2 - k_4 x_3,\\
    \dot{x}_4 = \frac{k_1x_1}{3(k_2 + x_1)} e^{-k_3 t} + \frac{1}{2}k_4 x_2 + k_4 x_3,\\
    y_1 = x_1,\\
    y_2 = x_2 + x_3,\\
    y_3 = x_4,
\end{cases}
\end{equation}
where $\bm{x} = (x_1, x_2, x_3, x_4)$, $\bm{y} = (y_1, y_2, y_3)$, $\bm{\mu} = (k_1, k_2, k_3, k_4)$, and $\bm{x}^\ast = (x_1^\ast, x_2^\ast, x_3^\ast, x_4^\ast)$.

The right-hand side of some of the equations involve an exponential function.
Let us denote $k_1 e^{-k_3 t}$ by $x_5$. 
By replacing all occurences of $k_1 e^{-k_3t}$ by $x_5$ and adding an extra equation $\dot{x}_5 = -k_3x_5$,  system~\eqref{eq:lip_orig} can be written using just rational functions  as follows
\begin{equation}\label{eq:lip_new}
\begin{cases}
    \dot{x}_1 = -\frac{x_1 x_5}{k_2  + x_1},\\
    \dot{x}_2 = \frac{2 x_1 x_5}{3(k_2 + x_1)} - k_4 x_2,\\
    \dot{x}_3 = \frac{1}{2} k_4 x_2 - k_4 x_3,\\
    \dot{x}_4 = \frac{x_1 x_5}{3(k_2 + x_1)} + \frac{1}{2}k_4 x_2 + k_4 x_3,\\
    \dot{x}_5 = -k_3 x_5,\\
    y_1 = x_1,\\
    y_2 = x_2 + x_3,\\
    y_3 = x_4,
\end{cases}
\end{equation}
where $\bm{x} = (x_1, x_2, x_3, x_4, x_5)$, $\bm{y} = (y_1, y_2, y_3)$, $\bm{\mu} = (k_2, k_3, k_4)$, and $\bm{x}^\ast = (x_1^\ast, x_2^\ast, x_3^\ast, x_4^\ast, k_1)$.

Models~\eqref{eq:lip_orig} and~\eqref{eq:lip_new} have the same set $\bm{\mu} \cup \bm{x}^\ast$ with the only difference that $k_1$ has been moved from $\bm{\mu}$ to $\bm{x}^\ast$.
One can see that in the sense of identifiability models~\eqref{eq:lip_orig} and~\eqref{eq:lip_new} are equivalent.
Similar change of variables is possible in many cases when the right-hand side of some of the equations is not a rational function (see also Section~\ref{sec:Goodwin}).

\paragraph{Results} All the parameters and initial conditions of~\eqref{eq:lip_new} (and, consequently, \eqref{eq:lip_orig}) are globally identifiable.

\paragraph{Source code:}
\url{https://github.com/pogudingleb/SIAN/blob/master/examples/Lipolysis.mpl}.

%%%%%%%%

\subsection{Goodwin oscillator}\label{sec:Goodwin}

\paragraph{Purpose of the example}
\begin{itemize}
    \item to show that locally but not globally identifiable parameters appear even in small systems arising in real-life systems;
    \item to show how one can handle the case 
    in which the right-hand side of some of the equations is not a rational function of the parameters.
\end{itemize}

\paragraph{System and discussion} The following model describes the oscillations in enzyme kinetics~\cite{Goodwin} and has been already used as a benchmark for software for identifiability analysis in~\cite[Case~1]{comparison}.
\begin{equation}\label{eq:Goodwin_orig}
  \begin{cases}
    \dot{x}_1 = -bx_1  + \frac{a}{A + x_3^\sigma},\\
    \dot{x}_2 = \alpha x_1 - \beta x_2,\\
    \dot{x}_3 = \gamma x_2 - \delta x_3,\\
    y_1 = x_1,
  \end{cases}
\end{equation}
where $\bm{x} = (x_1, x_2, x_3)$, $\bm{u} = \varnothing$, $\bm{y} = (y_1)$, $\bm{\mu} = (a, A, b, \alpha, \beta, \gamma, \delta, \sigma)$, $\bm{x}^\ast = (x_1^\ast, x_2^\ast, x_3^\ast)$.

To bring system~\eqref{eq:Goodwin_orig} to the form~\eqref{eq:main}, we introduce a new parameter $c$ and a new state variable $x_4$ defined by 
\begin{equation}\label{eq:Goodwin_change}
  c = \frac{A}{a}, \quad x_4 = \frac{x_3^\sigma}{a}.
\end{equation}
Then the first equation in~\eqref{eq:Goodwin_orig} can be rewritten as $\dot{x}_1 = -b x_1 + \frac{1}{c + x_4}$,
and an equation for $x_4$ can be derived as follows
\[
\dot{x_4} = \frac{1}{a}\sigma \dot{x}_3 x_3^{\sigma - 1} = \sigma \frac{x_3^\sigma}{a} \cdot \frac{\dot{x}_3}{x_3} = \sigma x_4 \frac{\gamma x_2 - \delta x_3}{x_3}.
\]
Thus, we can rewrite~\eqref{eq:Goodwin_orig} using just rational functions as
\begin{equation}\label{eq:Goodwin}
\begin{cases}
    \dot{x}_1 = -bx_1  + \frac{1}{c + x_4},\\
    \dot{x}_2 = \alpha x_1 - \beta x_2,\\
    \dot{x}_3 = \gamma x_2 - \delta x_3,\\
    \dot{x}_4 = \sigma x_4 \frac{\gamma x_2 - \delta x_3}{x_3},\\
    y_1 = x_1.
\end{cases}
\end{equation}
Here we have
\begin{itemize}
    \item $\bm{x} = (x_1, x_2, x_3, x_4)$,
    \item $\bm{u} = \varnothing$,
    \item $\bm{y} = (y_1)$,
    \item $\bm{\mu} = (b, c, \alpha, \beta, \gamma, \delta, \sigma)$,
    \item $\bm{x}^\ast = (x_1^\ast, x_2^\ast, x_3^\ast, x_4^\ast)$.
\end{itemize}

Our computations show that in the system~\eqref{eq:Goodwin}
\begin{itemize}
    \item $b, c, \sigma, x_1^\ast$, and $x_4^\ast$ are globally identifiable,
    \item $\beta$ and $\delta$ are locally but not globally identifiable,
    \item and $\alpha, \gamma, x_2^\ast$, and $x_3^\ast$ are non-identifiable.
\end{itemize}

Now we recall that, from~\eqref{eq:Goodwin_change}, we have 
\[
c = \frac{A}{a}\quad \text{and}\quad x_4^\ast = \frac{(x_3^\ast)^\sigma}{a}.
\]
If $a$ were locally identifiable, then the global identifiability of $x_4^\ast$ and $\sigma$ would imply that $x_3^\ast$ is locally identifiable.
Therefore, since $x_3^\ast$ is non-identifiable
, $a$ is non-identifiable.
Together with the global identifiability of $c$, the non-identifiability of $a$ yields the non-identifiability of $A$.
To sum up, the result of our identifiability analysis of~\eqref{eq:Goodwin_orig} is the following:
\begin{itemize}
    \item $b, \sigma$, and $x_1^\ast$ are globally identifiable,
    \item $\beta$ and $\delta$ are locally but not globally identifiable,
    \item and $a, A, \alpha, \gamma, x_2^\ast$, and $x_3^\ast$ are non-identifiable.
\end{itemize}

Once it is known that one of the parameters is not globally identifiable, one might want to understand where does the non-uniqueness come from and what to do about it.
Possible options include:
\begin{itemize}
    \item Among the possible parameter values, all but one violate some extra constraints coming from biology and can simply be  discarded at the data fitting stage.
    \item The non-uniqueness of the parameter value might arise from a flaw in the model that should be remedied by redesigning the model.
    \item The non-uniqueness has its own biological meaning, for example, it might indicate the existence of several distinct ``regimes'' of the model (see, for example, Section~\ref{sec:slowfast}).
    This biological meaning can be further used to identify the value of the parameter uniquely.
\end{itemize}

A natural step towards understanding the nature of the non-uniqueness of a parameter is to find a change of variables and parameters that leaves the outputs unchanged but changes the value of the parameter.
In the case of locally but not globally identifiable parameters $\beta$ and $\delta$ in~\eqref{eq:Goodwin}, one such change of variables and parameters is the following:

\begin{align}\label{eq:Goodwin_sym}
\begin{split}
  &x_1 \to x_1,\quad x_2 \to x_2 + \frac{\beta - \delta}{\gamma} x_3,\quad x_3 \to x_3,\quad x_4 \to x_4,\quad b \to b,\\ 
  &c \to c,\quad \alpha \to \alpha,\quad \beta \to \delta,\quad \gamma \to \gamma,\quad \delta \to \beta,\quad \sigma \to \sigma.
\end{split}
\end{align}

One can verify that~\eqref{eq:Goodwin_sym} preserves the output of~\eqref{eq:Goodwin} by a direct computation. 
Below we show one way to derive~\eqref{eq:Goodwin_sym} using our software.

\begin{enumerate}
    \item 
     From the intermediate results of the computation done by SIAN, we can extract that the pair of values $\{\beta, \delta\}$ is identifiable but it is impossible, based on the observations, to find out which of these two numbers is the value of $\beta$ and which one of them is the value of $\delta$.

    \item We try to find which of the state variables and/or parameters can be assumed to be known without making $\beta$ and $\delta$ globally identifiable.
    Using SIAN, one can verify (in a couple of seconds) that adding extra outputs $y_2 = x_3$, $y_3 = \alpha$, and $y_4 = \gamma$ does not make $\beta$ and $\gamma$ globally identifiable.

    \item Thus, there exists a change of variables and parameters that swaps $\beta$ and $\delta$ and leaves everything except for $\beta, \delta$, and $x_2$ unchanged.
    We can find the new function $\tilde{x}_2$ by looking at the third equation in~\eqref{eq:Goodwin} before and after the change of variables and parameters
    \begin{align*}
        \dot{x}_3 = \gamma x_2 - \delta x_3,\\
        \dot{x}_3 = \gamma \tilde{x_2} - \beta x_3.
    \end{align*}
    A direct computation shows that 
    \[
    \tilde{x}_2 = x_2 + \frac{\beta - \delta}{\gamma} x_3.
    \]
    Thus, we arrive at~\eqref{eq:Goodwin_sym}.
\end{enumerate}

\paragraph{Results}
\begin{itemize}
    \item $b, \sigma$, and $x_1^\ast$ are globally identifiable,
    \item $\beta$ and $\delta$ are locally but not globally identifiable,
    \item and $a, A, \alpha, \gamma, x_2^\ast$, and $x_3^\ast$ are non-identifiable.
\end{itemize}

\paragraph{Source code:}
 \url{https://github.com/pogudingleb/SIAN/blob/master/examples/Goodwin.mpl}.
    
%%%%%%%%%%%%%%

%%%%%%%%%%%%%%%%%%%%%%%%%%%%%%%%%%%

\section{Benchmarks}\label{sec:Benchmarks}

Section~4 of the paper compares performance of SIAN, GenSSI~2.0, COMBOS, and DAISY.
For the convenience of the reader, we reproduce the table with the runtimes (see Table~\ref{table:runtimes}).
The purpose of this section is to describe the used benchmark problems.
The source files of the benchmark problems for GenSSI 2.0, COMBOS, and DAISY are available in the Supplementary Data at~\url{https://cs.nyu.edu/~pogudin/SupplementaryData.zip}.

\begin{table}[h!]
\caption{Runtimes (in minutes) on benchmark problems}\label{table:runtimes}
\begin{center}
\begin{tabular}{|l|r|r|r|r|}
\hline
  Example & GenSSI~2.0 & COMBOS  & DAISY & SIAN \\
\hline
  Chemical Reaction (\ref{sec:chemical}) &$\ast$ & $\ast\ast$ & $>$ \hfill $6{,}000$ & $< \hspace{1mm}1$\\
  \hline
  HIV (\ref{sec:HIV}) & $>$ \hspace{1mm} $12{,}000$ & $\ast\ast$ & $>$ \hfill $6{,}600$ & $< \hspace{1mm}1$ \\
  \hline
  SIRS w/ forcing (\ref{sec:SIRS}) & $>$ \hspace{1mm} $12{,}000$ & $\ast\ast$ & $>$ \hfill $6{,}600$ & $< \hspace{1mm}1$ \\
  \hline
  Cholera (\ref{sec:Cholera}) &$\ast$ & $85\;$ & $30$  & $3$\\
  \hline
  Protein complex (\ref{sec:NFkB}) & $>$ \hspace{1mm} $12{,}000$ &$\ast\ast$ & $>$ \hfill $6{,}600$ & $47$\\
\hline
  Pharmacokinetics (\ref{sec:Pharm}) & $>$ \hspace{1mm} $12{,}000$ &$\ast\ast$  & $>$ \hfill $7{,}800$ & $962$\\
\hline
\end{tabular}
\end{center}
\small
$\ast$: \;\;\;\;\;GenSSI 2.0 returns~``{\it Warning: Unable to find explicit solution.}''\\
$\ast\ast$:\;\;\; COMBOS \ returns~``{\it Model may have been entered incorrectly or cannot be solved with COMBOS algorithms.}''\\
\end{table}

All the results presented in the rest of the section are computed with probability of correctness $p = 0.99$.

\subsection{Chemical Reaction}\label{sec:chemical}

\paragraph{System} The following system of ODEs corresponds to a chemical reaction network~\cite[Eq.~3.4]{ConradiShiu},  which is a reduced fully processive, $n$-site phosphorylation network.
\begin{align*}
\begin{cases}
\dot x_1 & = -\mu_1 x_1 x_2 + \mu_2 x_4 + \mu_4 x_6, \\ 
\dot x_2 & = -\mu_1 x_1 x_2 + \mu_2 x_4 + \mu_3 x_4, \\ 
\dot x_3 & = \mu_3 x_4 + \mu_5 x_6 - \mu_6 x_3 x_5, \\ 
\dot x_4 & = \mu_1 x_1 x_2 - \mu_2 x_4 - \mu_3 x_4, \\ 
\dot x_5 & = \mu_4 x_6 + \mu_5 x_6 - \mu_6 x_3 x_5, \\ 
\dot x_6 & = -\mu_4 x_6 - \mu_5 x_6 + \mu_6 x_3 x_5,\\
y_1 & = x_2,\\
y_2 & = x_3
\end{cases}%
\end{align*}
Here we have
\begin{itemize}
    \item $\bm{x} = (x_1, x_2, x_3, x_4, x_5, x_6)$,
    \item $\bm{u} = \varnothing$, 
    \item $\bm{y} = (y_1, y_2)$,
    \item $\bm{\mu} = (\mu_1, \mu_2, \mu_3, \mu_4, \mu_5, \mu_6)$,
    \item $\bm{x}^{\ast} = (x_1^\ast, x_2^\ast, x_3^\ast, x_4^\ast, x_5^\ast, x_6^\ast)$.
\end{itemize} 

\paragraph{Source code}
\begin{itemize}
    \item {SIAN}: \url{https://github.com/pogudingleb/SIAN/blob/master/examples/ChemicalReactionNetwork.mpl}.
    \item DAISY: file \emph{DAISY/ChemicalReactionNetwork.txt} in the \href{https://cs.nyu.edu/~pogudin/SupplementaryData.zip}{Supplementary Data}.
    \item COMBOS: file \emph{COMBOS/ChemicalReactionNetwork.txt} in the \href{https://cs.nyu.edu/~pogudin/SupplementaryData.zip}{Supplementary Data}.
    \item GenSSI~2.0: file \emph{GenSSI2/CRN.m} in the \href{https://cs.nyu.edu/~pogudin/SupplementaryData.zip}{Supplementary Data}.
\end{itemize}

\paragraph{Result} All the parameters $\bm{\mu}$ and initial conditions $\bm{x}^\ast$ are globally identifiable.

%%%

\subsection{HIV}\label{sec:HIV}

\paragraph{System} 
Consider the following model of HIV~\citep[Equation~(6)]{WN02} that describes immune impairment dynamics. 
\[
\begin{cases}
\dot{x} &= \lambda - d x - \beta x v,\\
\dot{y} &= \beta x v - a y,\\
\dot{v} &= k y - u v,\\
\dot{w} &= c z y w - c * q * y w - b w,\\
\dot{z} &= c q y w - h z,\\
y_1 &= w,\\
y_2 &= z.
\end{cases}
\]
Here we have
\begin{itemize}
    \item $\bm{x} = (x, y, v, w, z)$,
    \item $\bm{u} = \varnothing$,
    \item $\bm{y} = (y_1, y_2)$,
    \item $\bm{\mu} = (\beta, \lambda, a, c, d, h, k, q, u)$,
    \item $\bm{x}^\ast = (x^\ast, y^\ast, v^\ast, w^\ast, z^\ast)$.
\end{itemize}

\paragraph{Source code}
\begin{itemize}
    \item {SIAN}: \url{https://github.com/pogudingleb/SIAN/blob/master/examples/HIV2.mpl}.
    \item DAISY: file \emph{DAISY/HIV2.txt} in the \href{https://cs.nyu.edu/~pogudin/SupplementaryData.zip}{Supplementary Data}.
    \item COMBOS: file \emph{COMBOS/HIV2.txt} in the \href{https://cs.nyu.edu/~pogudin/SupplementaryData.zip}{Supplementary Data}.
    \item GenSSI~2.0: file \emph{GenSSI2/HIV2.m} in the \href{https://cs.nyu.edu/~pogudin/SupplementaryData.zip}{Supplementary Data}.
\end{itemize}

\paragraph{Results} $a, b, d, h, q, u, w^\ast$ and $z^\ast$ are globally identifiable, $\beta, \lambda, c, k, x^\ast, y^\ast$ and $v^\ast$ are non-identifiable.

%%%

\subsection{SIRS with forcing}\label{sec:SIRS}

\paragraph{System} The following model is an extension of the SIRS model that incorporates the seasonal nature of transmission of RSV~\citep[Equations~(7-11)]{CML2009}.
\[
\begin{cases}
\dot{s} &= \mu - \mu s - b_0 (1 + b_1 x_1) i s + g r,\\
\dot{i} &= b_0(1 + b_1 x_1) i s - (\nu + \mu) i,\\
\dot{r} &= \nu i- (\mu + g) r,\\
\dot{x_1} &= -M x_2,\\
\dot{x_2} &= M x_1,\\
y_1 &= i,\\
y_2 &= r.
\end{cases}
\]
Here we have
\begin{itemize}
    \item $\bm{x} = (s, i, r, x_1, x_2)$,
    \item $\bm{u} = \varnothing$,
    \item $\bm{y} = (y_1, y_2)$,
    \item $\bm{\mu} = (\mu, \nu, b_0, b_1, g, M)$,
    \item $\bm{x}^\ast = (s^\ast, i^\ast, r^\ast, x_1^\ast, x_2^\ast)$.
\end{itemize}

\paragraph{Source code}
\begin{itemize}
    \item {SIAN}: \url{https://github.com/pogudingleb/SIAN/blob/master/examples/SIRSForced.mpl}.
    \item DAISY: file \emph{DAISY/SIRSForced.txt} in the \href{https://cs.nyu.edu/~pogudin/SupplementaryData.zip}{Supplementary Data}.
    \item COMBOS: file \emph{COMBOS/SIRSForced.txt} in the \href{https://cs.nyu.edu/~pogudin/SupplementaryData.zip}{Supplementary Data}.
    \item GenSSI~2.0: file \emph{GenSSI2/SIRSForced.m} in the \href{https://cs.nyu.edu/~pogudin/SupplementaryData.zip}{Supplementary Data}.
\end{itemize}

\paragraph{Results} $b_0, g, \mu, \nu, s^\ast, i^\ast, r^\ast$ are globally identifiable, $M$ is locally identifiable, but not globally identifiable, and $b_1, x_1^\ast, x_2^\ast$ are non-identifiable.

%%%

\subsection{Cholera}\label{sec:Cholera}

\paragraph{System}
The following version of SIWR is an extension of the SIR
model, see~\cite[Eq.~3]{Cholera}:  
\[
\begin{cases}
\dot{s} & = \mu - \beta_I s i - \beta_W s w - \mu s + \alpha r, \\ 
\dot{i} & = \beta_W s w + \beta_I s i - \gamma i - \mu i, \\ 
\dot{w} & = \xi (i - w), \\ 
\dot{r} & = \gamma i - \mu r - \alpha r,\\
y_1 &= \kappa i,\\
y_2 &= s + i + r,
\end{cases}%
\]
where $s$, $i$, and $r$ stand for the fractions of the population that are
susceptible, infectious, and recovered, respectively.  The variable $w$
represents the concentration of the bacteria in the environment.
Here we have
\begin{itemize}
    \item $\bm{x} = (s, i, w, r)$,
    \item $\bm{u} = \varnothing$,
    \item $\bm{y} = (y_1, y_2)$,
    \item $\bm{\mu} = (\mu, \beta_I, \beta_W, \alpha, \gamma, \xi, \kappa)$,
    \item $\bm{x}^\ast = (s^\ast, i^{\ast}, w^{\ast}, r^{\ast})$.
\end{itemize}

\paragraph{Source code}
\begin{itemize}
    \item {SIAN}: \url{https://github.com/pogudingleb/SIAN/blob/master/examples/Cholera.mpl}.
    \item DAISY: file \emph{DAISY/Cholera.txt} in the \href{https://cs.nyu.edu/~pogudin/SupplementaryData.zip}{Supplementary Data}.
    \item COMBOS: file \emph{COMBOS/Cholera.txt} in the \href{https://cs.nyu.edu/~pogudin/SupplementaryData.zip}{Supplementary Data}.
    \item GenSSI~2.0: file \emph{GenSSI2/Cholera.m} in the \href{https://cs.nyu.edu/~pogudin/SupplementaryData.zip}{Supplementary Data}.
\end{itemize}

\paragraph{Result} All the parameters $\bm{\mu}$ and initial conditions $\bm{x}^\ast$ are globally identifiable.

%%%

\subsection{Protein Complex (NF$\kappa$B)}\label{sec:NFkB}

  Consider the model of NF$\kappa$B regulatory module proposed in~\citep{NFkB_first}
  (see also~\citep{NFkB_further} and~\citep[Case~6]{comparison}) defined by 
  the following system~\citep[Equation~27]{comparison}
  \begin{equation}
  \begin{cases}
    \dot x_1 = k_{prod} - k_{deg} x_1 - k_1 x_1  u,\\
    \dot x_2 = -k_3 x_2 - k_{deg} x_2 - a_2 x_2 x_{10} + t_1 x_4 - a_3 x_2 x_{13} + t_2 x_5 + (k_1 x_1 - k_2 x_2 x_8) u,\\
    \dot x_3 = k_3 x_2 - k_{deg} x_3 + k_2 x_2 x_8 u,\\
    \dot x_4 = a_2 x_2 x_{10} - t_1 x_4,\\
    \dot x_5 = a_3 x_2 x_{13} - t_2 x_5,\\
    \dot x_6 = c_{6a} x_{13} - a_1 x_6 x_{10} + t_2 x_5 - i_1 x_6,\\
    \dot x_7 = i_1 k_v x_6 - a_1 x_{11} x_7,\\
    \dot x_8 = c_4 x_9 - c_5 x_8,\\
    \dot x_9 = c_2 + c_1 x_7 - c_3 x_9,\\
    \dot x_{10} = -a_2 x_2  x_{10} - a_1 x_{10} x_6 + c_{4a} x_{12} - c_{5a} x_{10} - i_{1a} x_{10} + e_{1a} x_{11},\\
    \dot x_{11} = -a_1 x_{11} x_7 + i_{1a} k_v x_{10} - e_{1a} k_v x_{11},\\
    \dot x_{12} = c_{2a} + c_{1a} x_7 - c_{3a} x_{12},\\
    \dot x_{13} = a_1 x_{10} x_6 - c_{6a} x_{13} - a_3 x_2 x_{13} + e_{2a} x_{14},\\
    \dot x_{14} = a_1 x_{11} x_7 - e_{2a} k_v x_{14},\\
    \dot x_{15} = c_{2c} + c_{1c} x_7 - c_{3c} x_{15},\\
    y_1 = x_2,\\ 
    y_2 = x_{10} + x_{13},\\ 
    y_3 = x_9,\\
    y_4 = x_1 + x_2 + x_3,\\
    y_5 = x_7,\\
    y_6 = x_{12}
  \end{cases}  
  \end{equation}
  The values of all the parameters except $t_1, t_2, c_{3a}, c_{4a}, c_5, k_1, k_2, k_3, k_{prod}, k_{deg}, i_1, e_{2a}, i_{1a}$ are known from the existing literature (see~\citep[Table~1]{NFkB_further}).
  Here we have
  \begin{itemize}
      \item $\mathbf{x} = (x_1, x_2, \ldots, x_{15})$,
      \item $\mathbf{u} = (u)$,
      \item $\mathbf{y} = (y_1, y_2, \ldots, y_6)$,
      \item $\mathbf{\mu} = t_1, t_2, c_{3a}, c_{4a}, c_5, k_1, k_2, k_3, k_{prod}, k_{deg}, i_1, e_{2a}, i_{1a}$,
      \item $\mathbf{x}^\ast = (x_1^\ast, x_2^\ast, \ldots, x_{15}^\ast)$.
  \end{itemize}

\paragraph{Source code}
\begin{itemize}
    \item {SIAN}: \url{https://github.com/pogudingleb/SIAN/blob/master/examples/NFkB.mpl}.
    \item DAISY: file \emph{DAISY/NFkB.txt} in the \href{https://cs.nyu.edu/~pogudin/SupplementaryData.zip}{Supplementary Data}.
    \item COMBOS: file \emph{COMBOS/NFkB.txt} in the \href{https://cs.nyu.edu/~pogudin/SupplementaryData.zip}{Supplementary Data}.
    \item GenSSI~2.0: file \emph{GenSSI2/NFkB.m} in the \href{https://cs.nyu.edu/~pogudin/SupplementaryData.zip}{Supplementary Data}.
\end{itemize}

\paragraph{Result} All the parameters $\bm{\mu}$ and initial values $\bm{x}^\ast$ except $x_{15}^\ast$ are globally identifiable.
$x_{15}^\ast$ is non-identifiable.

%%%%%%%%%%%%%%

\subsection{Pharmacokinetics}\label{sec:Pharm}

\paragraph{System}
This is a simplified version of a model arising in pharmacokinetics~\citep{Pharm}:
  \begin{equation}\label{eq:Pharm}
  \begin{cases}
    \dot x_1 = a (x_2 - x_1) - \frac{k_a V_m x_1}{k_c k_a + k_c x_3 + k_a x_1},\\
    \dot x_2 = a (x_1 - x_2),\\
    \dot x_3 = b_1(x_4 - x_3) - \frac{k_cV_m x_3}{k_c k_a + k_c x_3 + k_a x_1},\\
    \dot x_4 = b_2(x_3 - x_4),\\
    y = x_1.
  \end{cases}  
  \end{equation}
Here we have
\begin{itemize}
    \item $\bm{x} = (x_1, x_2, x_3, x_4)$,
    \item $\bm{u} = \varnothing$,
    \item $\bm{y} = (y)$,
    \item $\bm{\mu} = (a, b_1, b_2, k_a, k_c, V_m)$,
    \item $\bm{x}^\ast = (x_1^\ast, x_2^\ast, x_3^\ast, x_4^\ast)$.
\end{itemize}

\paragraph{Source code}
\begin{itemize}
    \item {SIAN}: \url{https://github.com/pogudingleb/SIAN/blob/master/examples/Pharm.mpl}.
    \item DAISY: file \emph{DAISY/Pharm.txt} in the \href{https://cs.nyu.edu/~pogudin/SupplementaryData.zip}{Supplementary Data}.
    \item COMBOS: file \emph{COMBOS/Pharm.txt} in the \href{https://cs.nyu.edu/~pogudin/SupplementaryData.zip}{Supplementary Data}.
    \item GenSSI~2.0: file \emph{GenSSI2/Pharmacokinetics.m} in the \href{https://cs.nyu.edu/~pogudin/SupplementaryData.zip}{Supplementary Data}.
\end{itemize}

\paragraph{Result} All the parameters $\bm{\mu}$ and initial values $\bm{x}^\ast$ are globally identifiable.

%%%%%%%%%%%%%%%%%%%

%%%%%%%%%%%%%%%%%%%

\bibliographystyle{abbrvnat}
\bibliography{bibdata}